\documentclass[twocolumn,twoside]{Jornadas}
\usepackage[utf8]{inputenc}
\usepackage[spanish]{babel}
\usepackage{listings}
\usepackage{algorithm}
\usepackage{algorithmicx}
\usepackage{algcompatible}
\usepackage{adjustbox}
\usepackage{graphicx}
\usepackage{color}
\usepackage{caption}
\usepackage[caption=false,font=footnotesize]{subfig}
\usepackage{placeins}
\usepackage{hyperref}
\usepackage{url}

\usepackage{amsmath}
\usepackage{tikz}
\usepackage{mathdots}
\usepackage{yhmath}
\usepackage{cancel}
\usepackage{color}
\usepackage{array}
\usepackage{multirow}
\usepackage{amssymb}
\usepackage{tabularx}
\usepackage{extarrows}
\usepackage{booktabs}
\usetikzlibrary{fadings}
\usetikzlibrary{patterns}
\usetikzlibrary{shadows.blur}
\usetikzlibrary{shapes}
\usepackage{cite}
\usepackage{enumitem}
\usepackage{comment}
\usepackage{xspace}

\definecolor{gray97}{gray}{.97}
\definecolor{gray75}{gray}{.75}
\definecolor{gray45}{gray}{.45}

\def\BibTeX{{\rm B\kern-.05em{\sc i\kern-.025em b}\kern-.08em
    T\kern-.1667em\lower.7ex\hbox{E}\kern-.125emX}}

\newcommand{\bash}{\textsc{Bash}\xspace}

\begin{document}

%\title{Desarrollo de un concurso de programación para la enseñanza de Bash}
\title{Motivando el uso y aprendizaje de Bash a través de concursos de programación}

\author{%
     Luis Costero, Jorge Villarrubia y Francisco D. Igual%
     \thanks{Dpto. de Arquitectura de Computadores y Automática. Universidad Complutense de Madrid, e-mail: {\tt \{lcostero,jorvil01,figual\}@ucm.es}}
}

\maketitle
% Oculta las cabeceras y los números de página.
% Ambos elementos se añadirán durante la edición de las actas completas.
\markboth{}{}
\pagestyle{empty} 
\thispagestyle{empty} % Oculta el número de la primera página

\begin{abstract}
El aprendizaje de la línea de comandos y el uso de Bash son competencias fundamentales en entornos de administración de sistemas, desarrollo de software y ciencia de datos. Sin embargo, su enseñanza ha sido relegada en muchos planes de estudio, a pesar de su relevancia en el ámbito profesional. Para abordar esta carencia, hemos desarrollado un concurso interactivo que incentiva a los estudiantes a mejorar sus habilidades en Bash a través de desafíos prácticos y competitivos. Este enfoque gamificado busca motivar el aprendizaje autónomo y reforzar el dominio de la línea de comandos en un contexto dinámico. Los resultados han sido prometedores: de los 26 estudiantes participantes, el 85\% consideró que la actividad fue útil para mejorar sus conocimientos y un 71\% manifestó la necesidad de profundizar más en Bash de cara al futuro académico y profesional. Estos hallazgos sugieren que este tipo de iniciativas pueden ser una estrategia efectiva para fomentar el aprendizaje de Bash en entornos académicos.
\end{abstract}

\begin{keywords}
Docencia, Bash, Gamificación, Sistemas de corrección automática.
\end{keywords}

%Añade los ficheros que necesites:

%!TEX root = main.tex
\section{Introducción}\label{sec:intro}

%%INTRO DE BASH
\PARstart{E}{n} el ámbito universitario, el aprendizaje de herramientas fundamentales para la administración de sistemas y la automatización de tareas es esencial para los estudiantes, tanto para su éxito académico como para mejorar sus perspectivas laborales en la industria tecnológica.
\bash ({\bf B}ourne-{\bf a}gain {\bf sh}ell), uno de los intérpretes de comandos más utilizados en entornos Unix/Linux, y una herramienta clave para desarrolladores y administradores de sistemas. Sin embargo, su enseñanza no siempre forma parte del currículo académico de muchas universidades, debido a multitud de factores como pueden ser un enfoque centrado en lenguajes de programación más populares, la percepción de \bash como una herramienta de nicho o la omnipresencia de las interfaces de usuario (GUI) que ha reducido las oportunidades de interactuar con la computadora mediante línea de comandos~\cite{no-gui}.
Además, con frecuencia se asume que los alumnos aprenderán estas herramientas por su cuenta o existen dificultades para añadir cursos específicos debido a la carga curricular y limitaciones de tiempo~\cite{teaching_bash}.

No obstante, el dominio de \bash, y de otras utilidades de línea de comandos no solo es valioso en el ámbito profesional, sino que también beneficia a los estudiantes durante su formación académica.
Integrar \bash en el currículo puede potenciar tanto el aprendizaje como la empleabilidad: el trabajo en la línea de comandos brinda una interacción más directa con el sistema, obligando a comprender sus procesos subyacentes~\cite{no-gui}. Además, la capacidad de automatizar tareas mediante {\em scripts} incrementa la eficiencia en la resolución de problemas y en el manejo de sistemas, una competencia muy valorada en entornos profesionales~\cite{teaching_bash}.
El uso eficiente de la terminal permite, entre otras cosas, la automatización de tareas repetitivas, la gestión eficaz de archivos o un procesamiento de datos más ágil. Esto resulta especialmente útil en asignaturas que requieren manipulación de grandes volúmenes de información, ejecución de {\em scripts} o administración de entornos de desarrollo. Por ejemplo, en bioinformática, la programación en \bash es fundamental para procesar y analizar datos biológicos. Sin esta habilidad, los alumnos pueden enfrentarse a procesos manuales ineficientes que ralentizan su aprendizaje y dificultan la resolución de problemas en entornos reales; en ámbitos como la computación de altas prestaciones, la automatización de experimentos, así como la recogia y análisis de resultados, resulta clave de cara a aumentar la eficiencia del desarrollador.

%% INTRO DE GAMIFICACIÓN

La gamificación --uso de elementos y dinámicas de juego en entornos educativos-- se ha mostrado eficaz para aumentar la motivación, la retención y el compromiso de los alumnos, según numerosos estudios en educación superior, incluida la Informática~\cite{gamificacion-gen-z}. Este enfoque se basa en principios como la retroalimentación inmediata, la progresión a través de niveles, la obtención de recompensas y la competencia amistosa.
%A nivel general, el proceso de gamificación en la educación sigue varias etapas: primero, se definen los objetivos de aprendizaje; luego, se identifican las mecánicas de juego más adecuadas (puntos, insignias, tablas de clasificación, desafíos, etc.); después, se implementa el sistema gamificado y, finalmente, se evalúa su impacto en el rendimiento de los estudiantes. Estas dinámicas permiten transformar tareas que podrían percibirse como tediosas en experiencias interactivas y estimulantes.

Dentro de la gamificación, incorporar competiciones o concursos en la enseñanza, es una metodología particularmente efectiva para el aprendizaje de habilidades técnicas, como \bash. Estas metodologías explotan la naturaleza competitiva de los estudiantes y su deseo de logro, generando un entorno de aprendizaje más motivador. 
Al estructurar el aprendizaje en torno a concursos, los estudiantes enfrentan retos progresivos que requieren la aplicación práctica de sus conocimientos en tiempo real. Elementos como la resolución de problemas contrarreloj, el desbloqueo de niveles más avanzados a medida que se superan desafíos y la posibilidad de competir contra compañeros o equipos fomentan la participación activa y el desarrollo del pensamiento crítico. Además, los concursos proporcionan un entorno en el que el error se convierte en una oportunidad de aprendizaje, ya que los participantes reciben retroalimentación inmediata sobre su desempeño y pueden ajustar sus estrategias en función de los resultados obtenidos. Este enfoque no solo refuerza la comprensión de conceptos clave, sino que también promueve la autonomía y la resiliencia en la resolución de problemas, competencias esenciales tanto en el ámbito académico como en el profesional.
En particular, incorporar competencias o concursos en la enseñanza de programación explota la naturaleza competitiva de los estudiantes y su deseo de logro, generando un entorno de aprendizaje motivador. Estudios recientes señalan que esta estrategia gamificada con concursos conlleva notables mejoras en la motivación y el rendimiento del alumnado: por ejemplo, la introducción de competiciones de programación en cursos universitarios produjo un aumento significativo en la motivación y aprendizaje de los estudiantes~\cite{programming-contests}. Asimismo, muchos participantes reportan que competir les impulsa a profundizar más en los contenidos y obtener mejores resultados en sus asignaturas de programación~\cite{competitions-motivation}. Estas dinámicas de competición no solo hacen el aprendizaje más ameno, sino que exponen a los alumnos a problemas reales bajo presión de tiempo, reforzando sus habilidades de resolución de problemas y la aplicación práctica de sus conocimientos~\cite{programming-contests}.

En este sentido, este artículo describe nuestra experiencia docente durante dos cursos académicos (2023-2024 y 2024-2025), con la puesta en marcha de un concurso de programación en \bash para los estudiantes de la Facultad de Informática de la Universidad Complutense de Madrid, así como las conclusiones obtenidas.

Más concretamente, este artículo describe:
\begin{itemize}
    \item El diseño y puesta en marcha de un sistema automático para la corrección de ejercicios programados en lenguaje \bash, y su uso en la organización de concursos en un entorno real.
    \item La integración de elementos de gamificación, como la progresión incremental, la clasificación en tiempo real y un sistema interactivo de pistas, para potenciar el aprendizaje práctico y la motivación de los estudiantes.
    \item Una evaluación metodológica y experimental de la experiencia docente, analizando de forma cuantitativa y cualitativa los resultados del concurso, para extraer conclusiones sobre la efectividad del concurso y proponer mejoras a futuro.
\end{itemize}

El resto del artículo está organizado como sigue: la Sección~\ref{sec:background} revisa el trabajo relacionado, la Sección~\ref{sec:proposal} describe los objetivos y la solución propuesta, la Sección~\ref{sec:experiments} explica el desarrollo práctico de esta experiencia docente, la Sección~\ref{sec:results} analiza los resultados y la Sección~\ref{sec:conclusions} establece las conclusiones.

\section{Trabajo relacionado}\label{sec:background}

La gamificación es una técnica muy utilizada para la enseñanza, integrando contenidos educativos con elementos lúdicos a través de ``juegos serios'' \cite{deterding2011game}. Entre los ejemplos más destacados se encuentran plataformas orientadas al aprendizaje de idiomas, como Duolingo \cite{DuolingoReview}, que utiliza recompensas diarias y niveles progresivos para fomentar una mejora constante y gradual en el dominio de una nueva lengua. Por otro lado, entornos interactivos de cuestionarios como Kahoot \cite{KahootReview} permiten a los estudiantes responder preguntas académicas desde sus dispositivos móviles, acumulando puntos en función de la exactitud y rapidez de sus respuestas. Estos puntos se reflejan en un ranking en tiempo real, que el profesor utiliza para incentivar una sana competencia entre los alumnos.

Plataformas asíncronas de evaluación automática como UVa Online Judge \cite{revilla2008UVa}, Codeforces \cite{mirzayanov2020codeforces}, Kattis \cite{Kattis} o AceptaElReto \cite{gomez2017aceptaelreto}, son ampliamente utilizadas en educación superior para entrenar habilidades algorítmicas mediante la resolución de problemas. Permiten a los estudiantes enviar el código con la solución de un reto de programación en cualquier momento, proporcionando una retroalimentación inmediata (por ejemplo, veredicto \guillemotleft correcto\guillemotright, \guillemotleft tiempo límite excedido\guillemotright, \guillemotleft error de compilación\guillemotright, etc.) Sin embargo, los lenguajes de scripting, en particular \bash, no están permitidos o son residuales, ya que su aplicación a problemas de algoritmia dista mucho de su verdadera utilidad en tareas de automatización y administración de servidores.

Combinando ambas opciones surgen concursos de programación como SWERC \cite{SWERC}, AdaByron \cite{Ada-Byron}, ProgramaMe \cite{Programame} o la Olimpiada de Informática \cite{OIE}, en que los participantes se enfrentan a una batería de problemas en un tiempo limitado, con clasificaciones que se actualizan en tiempo real. Para el despliegue y la gestión de estos eventos, se utilizan diversas herramientas de código abierto como Mooshak \cite{Leal2003Mooshak} o DOMjudge \cite{DOMjudge}, que además de la clasificación incluyen la gestión de las inscripciones o la evaluación automática de soluciones. La mayoría de estos concursos se orienta a problemas de algoritmia y estructuras de datos, pero existen iniciativas con otros enfoques. Un ejemplo son los concursos de programación paralela organizados por SARTECO \cite{concursoSARTECO}, centrados en el uso de tecnologías como CUDA, MPI y OpenMP.

Si bien los concursos de programación siguen mayoritariamente orientados a problemas algorítmicos en lenguajes como C++ o Java, existen diversas plataformas de corrección automática en línea dedicadas específicamente a retos de automatización y administración de sistemas mediante \bash. Un ejemplo destacado es CMD Challenge \cite{CMDChallenge}, con una serie de ejercicios diseñados para practicar las principales utilidades de \bash, organizados en niveles de dificultad progresiva que el usuario debe superar. Esta plataforma incorpora elementos de gamificación de manera atractiva, ya que cada nivel está asociado a un animal, cuyo emoji se ``desbloquea'' y colecciona al completarlo. Otras webs, como Exercism, presentan una mecánica similar para aprender distintos lenguajes de programación, y en particular incluyen una itinerario específico para \bash \cite{ExercismBash}.

Sin embargo, la mayoría de los sistemas de corrección automática, en particular aquellos empleados en concursos, se enfocan en problemas algorítmicos, dejando de lado otras competencias cada vez más demandadas, como la automatización y la administración de sistemas mediante lenguajes de scripting, entre ellos \bash. Estas habilidades son esenciales en áreas como DevOps o la ciberseguridad, donde la gestión y orquestación eficiente de sistemas complejos es fundamental. Por otro lado, las pocas plataformas recientemente desarrolladas para \bash operan en entornos abstractos o simplificados, alejados de flujos de trabajo realistas, lo que reduce la preparación de los estudiantes para enfrentar entornos profesionales. Tratando de paliar estas limitaciones, en nuestro concurso los participantes trabajan directamente por línea de comandos en un servidor GNU/Linux a través de una sesión SSH, utilizando herramientas nativas sin capas de abstracción innecesarias. Además, nuestro enfoque incorpora funcionalidades innovadoras como un sistema de pistas que ayuda a los participantes a superar posibles bloqueos.

\section{Solución Propuesta}\label{sec:proposal}

La principal motivación que nos llevó a crear este concurso de programación en \bash fue aumentar el interés de los alumnos por este lenguaje y el uso de la línea de comandos en general. Demostrar las capacidades que éstas tienen, y motivar a los alumnos para que las usen puede ayudarles no solo en asignaturas más técnicas de la carrera (como puede ser Sistemas Operativos), sino también en su futuro laboral.
Más concretamente, a la hora de plantear el concurso, se definieron los siguientes objetivos, que dieron forma final al sistema desarrollado y a la experiencia docente:

%\begin{enumerate}[leftmargin=10pt]
\begin{enumerate}[start=1,label={\bfseries O\arabic*:}]

    \item Los estudiantes deberán interactuar con un sistema lo más realista posible, evitando soluciones que requieran aplicaciones específicas o añadan capas de abstracción innecesarias. Este punto nos lleva a descartar los sistemas existentes, que añaden complejidad al entorno, y a desarrollar un sistema nuevo lo más liviano posible, donde los concursantes interactúen con el sistema directamente a través de una línea de comandos.
    
    \item La retroalimentación obtenida por los estudiantes deberá de ser inmediata, evitando la pérdida de motivación e interés. Esto nos obliga a que el sistema corrija de forma automática las soluciones enviadas, evitando la intervención del profesor.

    \item Buscamos que los concursantes menos experimentados consoliden los conocimientos básicos antes de enfrentarse a retos mayores, promoviendo la competitividad sin perder de vista el aprendizaje. A diferencia de otros concursos, que presentan todos los problemas a la vez sin indicar su dificultad, nuestro sistema los revelará secuencialmente según se vayan resolviendo, en orden creciente de complejidad. Los problemas comenzarán con ejercicios muy básicos sobre el lenguaje, avanzando progresivamente hasta desafíos sofisticados. Así, los participantes novatos ganarán confianza y no intentarán problemas complejos sin haber resuelto los más sencillos. Por su parte, los concursantes con más conocimiento son guiados a comenzar por los problemas que pueden resolver rápido, para después enfrentarse a retos a su medida.
    
    \item Evitar que un estudiante pierda la motivación y desista en caso de que no ser capaz de resolver un problema concreto. Para ello, se ha implementado un sistema en el que los estudiantes puedan pedir pistas específicas para el problema, con una pequeña penalización en su puntuación final que desincentiva el uso innecesario de las pistas. Otras soluciones como permitir que resuelvan más de un problema a la vez se consideraron, pero dado que los problemas están ordenados por dificultad, es muy probable que si un alumno tiene problemas con un problema específico, también los vaya a tener con problemas sucesivos.

    \item Fomentar la competición entre estudiantes, y por tanto la motivación. Para ello, habrá un sistema de clasificación en tiempo real donde todos los estudiantes puedan ver el estado del resto de participantes en cualquier momento, comparándose entre ellos y obteniendo información general del concurso (e.g., cantidad de soluciones correctas, fallidas y pistas pedidas para cada problema del concurso).
    
\end{enumerate}

De forma resumida, la presente experiencia docente se fundamenta en un sistema automático que permite la interacción de los estudiantes con un sistema GNU/Linux real a través de la terminal, la verificación automática de las soluciones, la solicitud de pistas, la progresión gradual en la dificultad de los problemas a medida que se resuelven, y la clasificación de estudiantes de forma pública en tiempo real.

\subsection{Interacción y validación automática de soluciones}

El objetivo primordial del concurso es fomentar la familiarización de los 
estudiantes con  \bash y, en general, con el uso de la línea de comandos. Con 
este propósito, el sistema ha sido implementado sobre un servidor ejecutando Debian GNU/Linux, 
con el que los participantes interactúan directamente mediante sesiones SSH. 
Esta metodología garantiza un entorno de trabajo realista, donde los 
estudiantes utilizan herramientas propias del sistema operativo, en contraste 
con otras experiencias en las que la evaluación se realiza en un entorno 
aislado y con herramientas externas.

Al iniciar sesión, los usuarios acceden a un conjunto de problemas 
organizados en ficheros de texto distribuidos en carpetas, cada una 
correspondiente a un nivel de dificultad. Cada vez que un problema es resuelto 
correctamente, una nueva carpeta se muestra en el directorio de trabajo del 
usuario con el siguiente problema a resolver.

Para validar las soluciones, los participantes disponen de comandos 
específicos con los que comprobar si su propuesta es correcta y, en caso 
afirmativo, avanzar al siguiente nivel. La evaluación de las soluciones se 
lleva a cabo mediante un conjunto de ficheros de entrada ocultos, comparando 
las respuestas obtenidas con las esperadas. Este proceso se realiza a través 
de un sistema controlado de ejecución, implementado mediante herramientas 
del sistema operativo como \texttt{sudo}, \texttt{chroot} y listas de acceso, 
entre otras. Toda interacción de los estudiantes con el sistema es almacenada tanto en una base de datos (resultados de la evaluación, usados para calcular la puntuación y ranking del concurso), como en ficheros de logs (resultados detallados de cada evaluación, usados por el profesor para hacer un análisis más detallado a posteriori). 

Adicionalmente, los estudiantes tienen acceso a una utilidad que les permite 
validar sus soluciones frente a un conjunto reducido de casos de prueba, 
conocidos públicamente. Esta validación no afecta a la puntuación final, 
pero les proporciona retroalimentación sobre el funcionamiento de sus 
soluciones. Asimismo, el sistema ofrece una funcionalidad para la solicitud de 
pistas específicas para cada problema.

\subsection{Descripción de los problemas}

Cada problema consiste en la implementación de un {\em script} en lenguaje  \bash 
que, utilizando las construcciones propias del lenguaje y las utilidades 
básicas del sistema, resuelva la tarea planteada. El {\em script} puede recibir 
datos de entrada a través de argumentos o mediante la entrada estándar, y 
debe mostrar el resultado en la salida estándar.

Para cada problema, se proporciona un ejemplo de ejecución con una entrada 
conocida. Además, existe un conjunto de entradas ocultas utilizadas para 
validar la correcta resolución del problema. La evaluación se lleva a cabo 
comparando la salida obtenida por el {\em script} con la salida esperada para cada 
una de las entradas ocultas. Un conjunto más amplio y diverso de casos de 
prueba permite una evaluación más precisa de las soluciones enviadas.

El listado~\ref{lst:problema_ejemplo} presenta un ejemplo de problema mostrado a los participantes, junto 
con su correspondiente ejemplo de ejecución. 

\begin{lstlisting}[%
%
label=lst:problema_ejemplo,
caption={Ejemplo de enunciado de problema propuesto}, 
%
inputencoding=utf8,
     extendedchars=true,
     backgroundcolor=\color{gray97},
     stringstyle=\ttfamily,
     showstringspaces = false,
     basicstyle=\scriptsize\ttfamily,
     linewidth=1.0\columnwidth,
     breakindent=0pt,
     breaklines=true,
     breakatwhitespace=true,
     numbers=none,
     xleftmargin=0pt
]
 ------ [ENUNCIADO] -------------------
Eres un administrador de sistemas avanzado y has implementado políticas que te permiten almacenar los intentos fallidos de inicio de sesión de todos los usuarios del sistema. Para determinar si estás siendo víctima de un ataque, e intentar averiguar con qué usuario está intentando atacarte, te piden que crees un script que analice qué usuarios son los que han intentado acceder más veces al sistema de forma fallida.

Crea un programa que analice los intentos de inicio de sesión fallidos.  El programa recibirá como argumento un número N y la ruta de un fichero de log que contienen los últimos accesos fallidos.

El fichero de log está compuesto por múltiples líneas con el siguiente formato:

admin    ssh:notty    93.144.87.93     Fri Nov 10 14:56 - 14:56  (00:00)

La salida será un listado de los usuarios que han intentado acceder al menos N veces (de forma fallida), precedido del número de intentos. La salida deberá estar ordenada en función del número de intentos. En caso de empate, se ordenará en forma alfabética inversa.

 ------ [INPUT] -----------------------

$ ./programa.sh 7 /contest/files/public/lastb/intentos_acceso.txt

 ------ [OUTPUT] -----------------------

41 root
18 pi
9 admin
8 NL5xUDpV2xRa
7 craft
 --------------------------------------
\end{lstlisting}

\subsection{Sistema de pistas}

Además del ejemplo de ejecución proporcionado, los estudiantes pueden 
solicitar pistas en cualquier momento. Las pistas son ayudas específicas para 
cada problema y su uso es limitado. Las pistas son otorgadas únicamente cuando el concursante las solicita, y se muestran en orden decreciente de importancia, evitando que un concursante bloqueado en un problema necesita solicitar todas las pistas disponibles para poder avanzar. El número de pistas solicitadas influye 
en la puntuación final del concursante. El listado~\ref{lst:pistas_ejemplo} 
presenta un ejemplo de pistas asociadas al problema mostrado en el 
listado~\ref{lst:problema_ejemplo}.

\begin{lstlisting}[ %
%
label=lst:pistas_ejemplo,
caption={Ejemplo de pistas propuestas asociadas al enunciado mostrado en el listado~\ref{lst:problema_ejemplo}}, 
%
     inputencoding=utf8,
     extendedchars=true,
     backgroundcolor=\color{gray97},
     stringstyle=\ttfamily,
     showstringspaces = false,
     basicstyle=\scriptsize\ttfamily,
     linewidth=1.0\columnwidth,
     breakindent=0pt,
     breaklines=true,
     numbers=none,
     xleftmargin=0pt
]
 ------ [PISTAS] -----------------------

[[ 1 ]]   Sort por defecto ordena de forma alfabética. Si queremos ordenar de forma numérica, puedes utilizar la opción -h o -n (comprueba el manual).
Para ordenar de forma inversa, puedes utilizar la opción -r

[[ 2 ]]   Para iterar sobre una variable que guarda múltiples líneas, puedes utilizar el siguiente esquema:

variable=$(.....)

while read -r variable2; do
        [...]
        #trabajar con $variable2
        [...]
done <<< "$variable"
 --------------------------------------

 \end{lstlisting}

\subsection{Sistema de clasificación}

El sistema ofrece una clasificación en tiempo real que se puede consultar a través de un servidor 
web. El orden de los concursantes viene determinado principalmente por el número de 
problemas resueltos. En caso de empate, se utiliza un sistema de puntuación 
basado en el tiempo total empleado por el concursante en resolver los problemas (es decir, el 
tiempo transcurrido desde el inicio del concurso hasta la resolución del último 
problema aceptado), que aumenta en función del número de intentos fallidos 
y del número de pistas solicitadas.

Toda esta información está disponible públicamente a través del servidor 
web, permitiendo que cada concursante conozca el estado de los demás 
participantes, así como extraer información adicional sobre los problemas, como la 
dificultad medida en función del número de intentos realizados o el número 
de pistas solicitadas.

\subsection{Descripción técnica}

\begin{figure}[t]
    \centering
    \includegraphics[width=0.85\columnwidth]{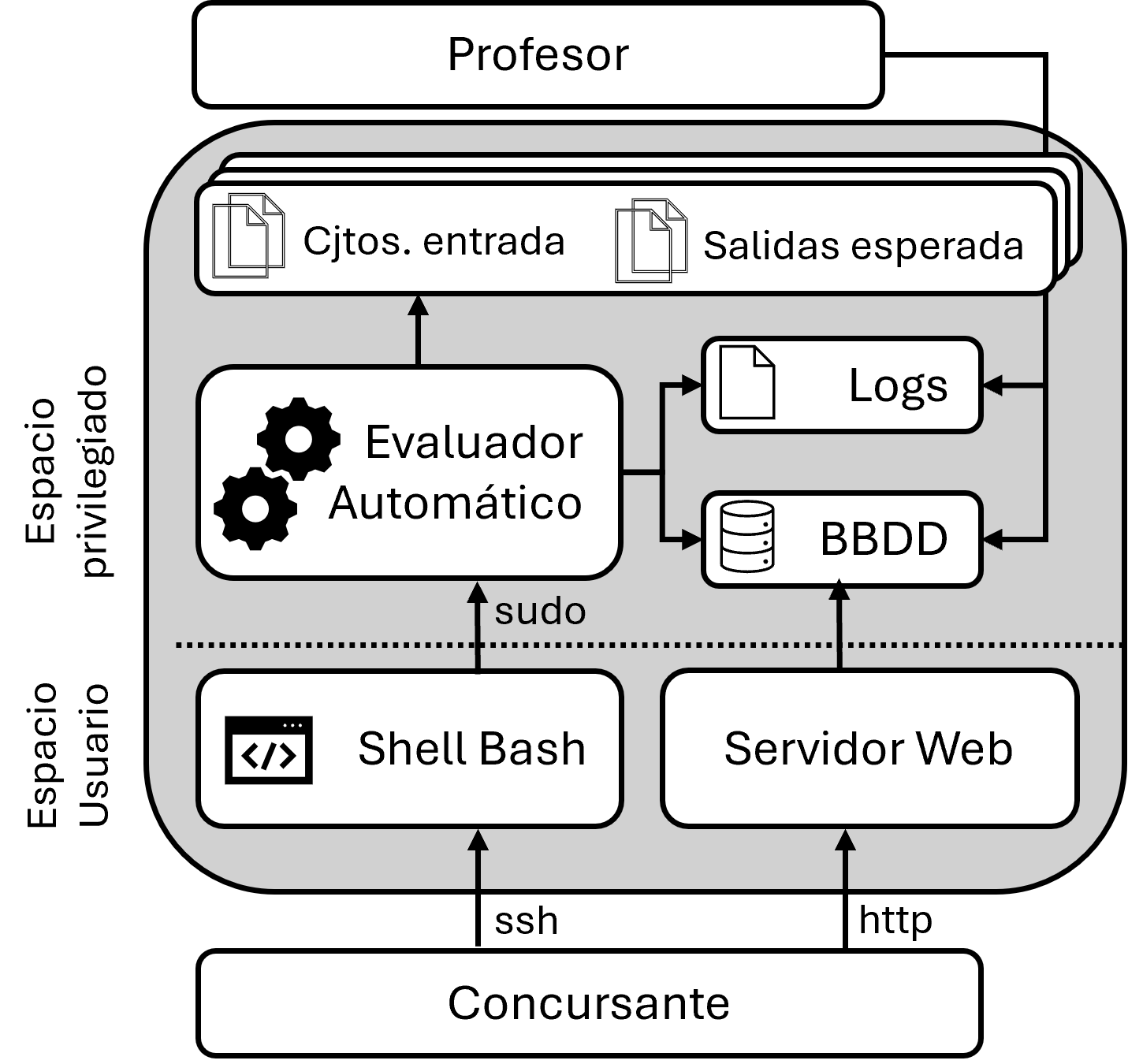}
    \caption{Esquema general del sistema propuesto}
    \label{fig:general-schema}
\end{figure}

La figura~\ref{fig:general-schema} muestra un esquema de los distintos componentes que forman el sistema. Como se ha comentado anteriormente, el sistema de evaluación está compuesto por una serie de elementos sobre un sistema Debian GNU/Linux, haciendo uso de las utilidades incluidas en este sistema operativo.
Para asegurar un concurso justo, y evitar posibles problemas durante el mismo, el sistema se ha desarrollado para ejecutarse en dos espacios distintos, denominados ``espacio de usuario'' y ``espacio privilegiado''. El primero con el que los concursantes interactúan a través de sesiones ssh con una terminal  \bash y a través de un servidor web, y el segundo en el que se realizan las evaluaciones de los envíos, y se almacena su resultado.
La separación entre espacios se realiza a través de distintos usuarios en el sistema, permisos de ficheros, y listas de control de acceso. La elevación de privilegios de un espacio al otro se hace a través de comandos específicos implementados dentro de la shell de usuario, que eventualmente acaban utilizando la utilidad \texttt{sudo} para elevar los privilegios.

Una vez el evaluador automático recibe un envío para su corrección, éste ejecuta el {\em script} enviado múltiples veces, cada vez con uno de los conjuntos de datos de entrada almacenado en el disco. La salida de cada ejecución es comparada con la salida esperada almacenada, y guardada en el fichero de log correspondiente por si es necesario realizar un análisis posterior. Además, el resultado conjunto de toda la evaluación es almacenado en la base de datos, y es utilizado para calcular y mostrar el ranking del concurso. Tanto el resultado final, como el resultado de cada uno de los casos de prueba pueden ser consultados por el profesor posteriormente.
La evaluación de las soluciones se hace en un entorno controlado a través de herramientas facilitadas por el sistema operativo (\texttt{chroot, cgroups, namespaces, ulimits, ...}).

Una vez un envío es correcto, el evaluador automático es responsable de actualizar la carpeta del usuario copiando el nuevo problema a resolver, y notificando al usuario cambiando su carpeta de trabajo (\texttt{cwd}).

\section{Desarrollo de la experiencia docente}\label{sec:experiments}

Hasta la fecha se han desarrollado dos ediciones del concurso descrito. Durante el año 2023-2024, 9 estudiantes participaron, aumentando el número hasta 27 en el curso 2024-2025. En ambos casos, la inscripción en el concurso era abierta a todos los estudiantes de la Facultad de Informática, y anunciada con un mes de antelación a través de los canales típicos de la facultad (pantallas en el hall, listas de correo, etc.). Dos de los estudiantes participaron en ambas ediciones del concurso. Esto es debido en parte a que, como se verá más adelante, la mayoría de concursantes pertenecen al último curso del grado, no pudiendo participar más de una edición consecutiva.
De forma adicional, a los alumnos matriculados en las asignaturas de Sistemas Operativos (tercer curso) y Ampliación de Sistemas Operativos y Redes (cuarto y quinto curso) se les mencionó explícitamente en clase, indicándoles que la nota final de la asignatura se vería incrementada en función del desempeño en el concurso (hasta dos puntos máximo sobre la parte práctica de la asignatura, lo que supone un máximo de 0.8 puntos extra en la nota final para el estudiante que resuelva todos los problemas). Un total de 34 alumnos participantes se encontraban matriculado en alguna de estas asignaturas.

Para preparar a los estudiantes de cara al concurso, se pidió a una de las asociaciones de estudiantes de la facultad la impartición de un curso de programación \bash (2 horas) unas semanas antes al desarrollo del concurso.

\begin{table*}[th]
    \centering
    \caption{Construcciones necesarias para resolver cada uno de los problemas propuestos}
    \setlength{\tabcolsep}{4pt}
    \begin{tabular}{ccccccccc}\toprule
    Problema && Aritmética & \shortstack{Filtrado\\(grep)} & Tuberías & \shortstack{Bucles y\\Condicionales} & \shortstack{Manipulación\\de texto} & \shortstack{Utilidades\\del sistema} & Ficheros \\\midrule
1  & & - & - & - & - & - & - & - \\
2  & & \checkmark & - & - & - & - & - & - \\
3  & & - & \checkmark & \checkmark & - & - & - & - \\
4  & & - & - & \checkmark & - & - & \checkmark & - \\
5  & & \checkmark & - & - & \checkmark & - & - & - \\
6  & & - & - & \checkmark & - & \checkmark & \checkmark & - \\
7  & & - & - & \checkmark & \checkmark & \checkmark & \checkmark & - \\
8  & & - & - & \checkmark & - & - & \checkmark & \checkmark \\
9  & & - & - & \checkmark & - & \checkmark & \checkmark & \checkmark \\
10 & & - & \checkmark & \checkmark & \checkmark & - & \checkmark & - \\
11 & & - & \checkmark & \checkmark & \checkmark & \checkmark & \checkmark & \checkmark \\\bottomrule

\end{tabular}

    \label{tab:descripcion_problemas}
\end{table*}

En cada edición del concurso se propusieron distintos problemas a resolver, intentando siempre cubrir todos los aspectos de programación de \bash. Por ejemplo, la Tabla~\ref{tab:descripcion_problemas} muestra los temas tratados por cada uno de los problemas propuestos en la edición 2024-2025. Como se ha descrito anteriormente, los problemas se muestran por orden de dificultad.

Durante el desarrollo del concurso, el primer problema se resolvió de forma guiada para mostrar la interacción con el sistema y explicar el funcionamiento del envío, solicitud de pistas y penalizaciones. Adicionalmente se proporcionó una guía resumida de las construcciones de \bash y comandos útiles del sistema. Aunque los participantes no tenían acceso a internet durante el concurso, sí que podían consultar las páginas del manual a través de la consola. La duración del concurso fue de 2 horas en ambas ediciones.

Para obtener un estudio más detallado del resultado de esta experiencia docente, se solicitó a los participantes que rellenaran una encuesta antes y después del concurso. En total 26 de los 36 participantes respondieron a las encuestas, cuyas preguntas se listan a continuación. La siguiente sección analiza los resultados de estas encuestas junto con los resultados objetivos obtenidos durante el concurso.

\subsection{Preguntas generales realizadas antes de iniciar el concurso}

Preguntas con resultados acotados {\small\em (1:nada, 2:poco, 3:medio, 4:alto, 5:muy alto)}:
\begin{enumerate}
    \item[A1] ¿Cuál consideras que es tu nivel de programación en shell/Bash?
    \item[A2] ¿Consideras que la programación en \mbox{shell/Bash} es una herramienta útil de cara a tu futuro académico o laboral?
    \item[A3] ¿Crees que el nivel de formación en programación en shell/Bash que se ofrece en las distintas asignaturas relacionadas del grado es suficiente?
    \item[A4] ¿Crees que sería conveniente intensificar o extender la formación en programación en shell/Bash en las asignaturas de grado?
    \item[A5] ¿Crees que sería conveniente intensificar o extender la formación en programación en shell/Bash mediante cursos o seminarios externos?
\end{enumerate}

Preguntas con resultados no numéricos (posibles respuestas entre paréntesis):
\begin{enumerate}
    \item[A6] ¿Cuál es tu principal motivación para participar en el concurso? {\small\emph{(Aprender más sobre \bash, Ganar experiencia en concursos, Interés general en la programación, Mejorar mis habilidades de resolución de problemas, Obtener reconocimiento académico, Divertirme y desafiarme a mí mismo/a)}}
    \item[A7] ¿Qué otros lenguajes de scripting dominas y/o usas frecuentemente? {\small\emph{(Python, Perl, Lua, Ruby, \mbox{Javascript}, Groovy)}}

\end{enumerate}

\subsection{Preguntas específicas realizadas al finalizar el concurso}

Preguntas con resultados acotados {\small\em (1:nada, 2:poco, 3:medio, 4:alto, 5:muy alto)}:
\begin{enumerate}
    \item[B1] ¿Crees que el concurso permite demostrar de forma correcta tu nivel de programación y resolución de problemas?
    \item[B2] ¿Consideras adecuada la recompensa que se obtiene tras la finalización del concurso en las asignaturas SO/ASOR?
\end{enumerate}

Preguntas con resultados no numéricos (posibles respuestas entre paréntesis):
\begin{enumerate}
    \item[B3] ¿Qué aspectos te han gustado más? {\small\emph{(\mbox{Hardware/software} disponible, Documentación aportada, Variedad de ejercicios, Nivel de dificultad de los ejercicios, Tiempo disponible)}}
    \item[B4] ¿Qué aspectos crees que podrían mejorarse de cara a próximas ediciones? {\small\emph{(Hardware/software disponible, Documentación aportada, Variedad de ejercicios, Nivel de dificultad de los ejercicios, Tiempo disponible, Nada)}}
\end{enumerate}

\section{Resultados experimentales}\label{sec:results}

Esta sección muestra un análisis objetivo de los resultados obtenidos durante el concurso~\ref{sec:resultados_concurso}, así como de las encuestas realizas a los alumnos antes del mismo (enfocadas a conocer la percepción de los estudiantes hacia \bash) en la sección~\ref{sec:encuestas_pre} y de las encuestas realizadas al finalizar el concurso (enfocadas a conocer la percepción de los estudiantes hacia el concurso) en la sección~\ref{sec:encuestas_post}.

\subsection{Sobre los resultados del concurso}\label{sec:resultados_concurso}

\begin{figure*}[t]
    \centering
    \subfloat[Información por problema\label{fig:resultados_problemas}]{%
        \includegraphics[width=0.49\textwidth]{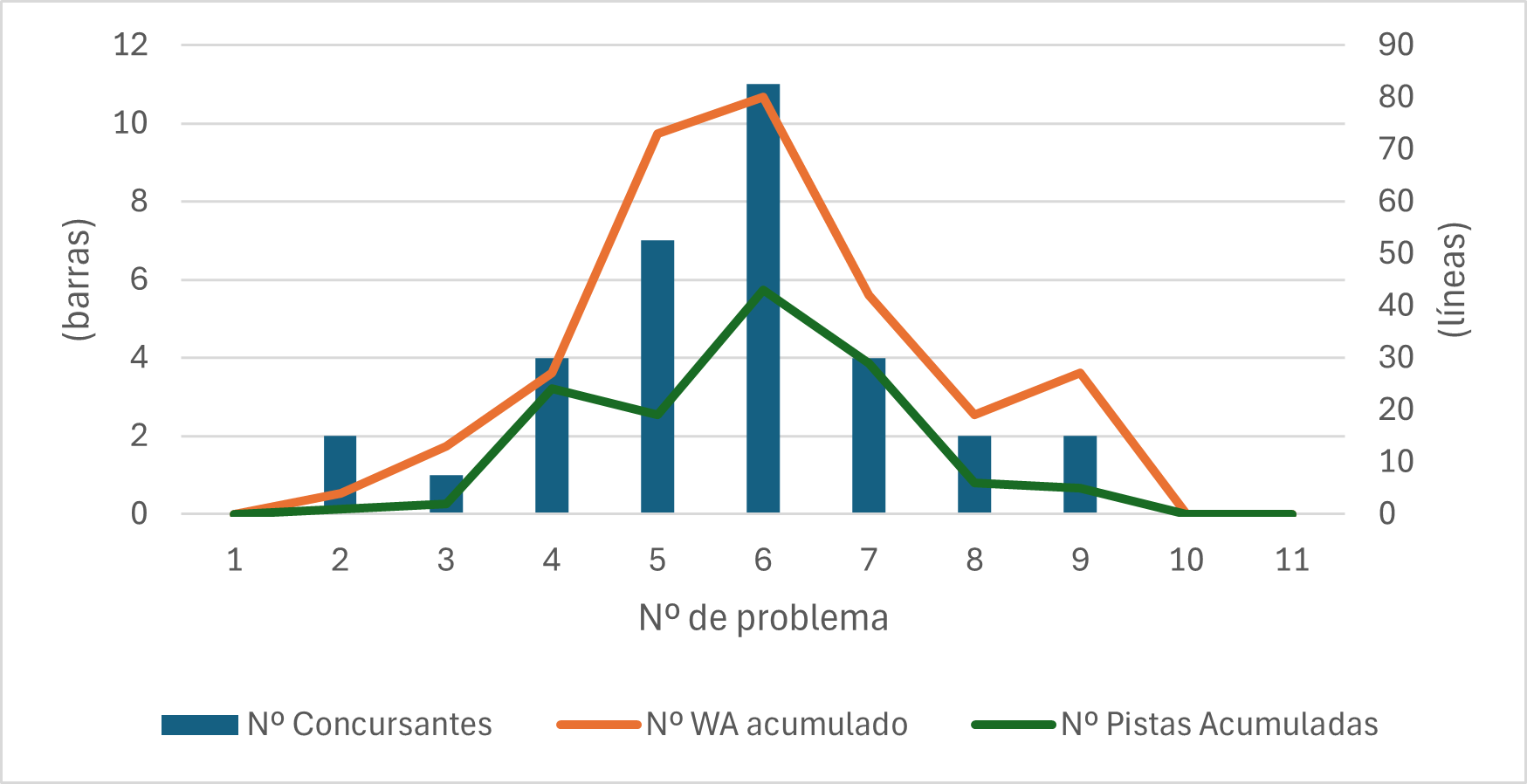}
    }
    \hfill
    \subfloat[Información por curso matriculado\label{fig:resultado_curso}]{%
        \includegraphics[width=0.49\textwidth]{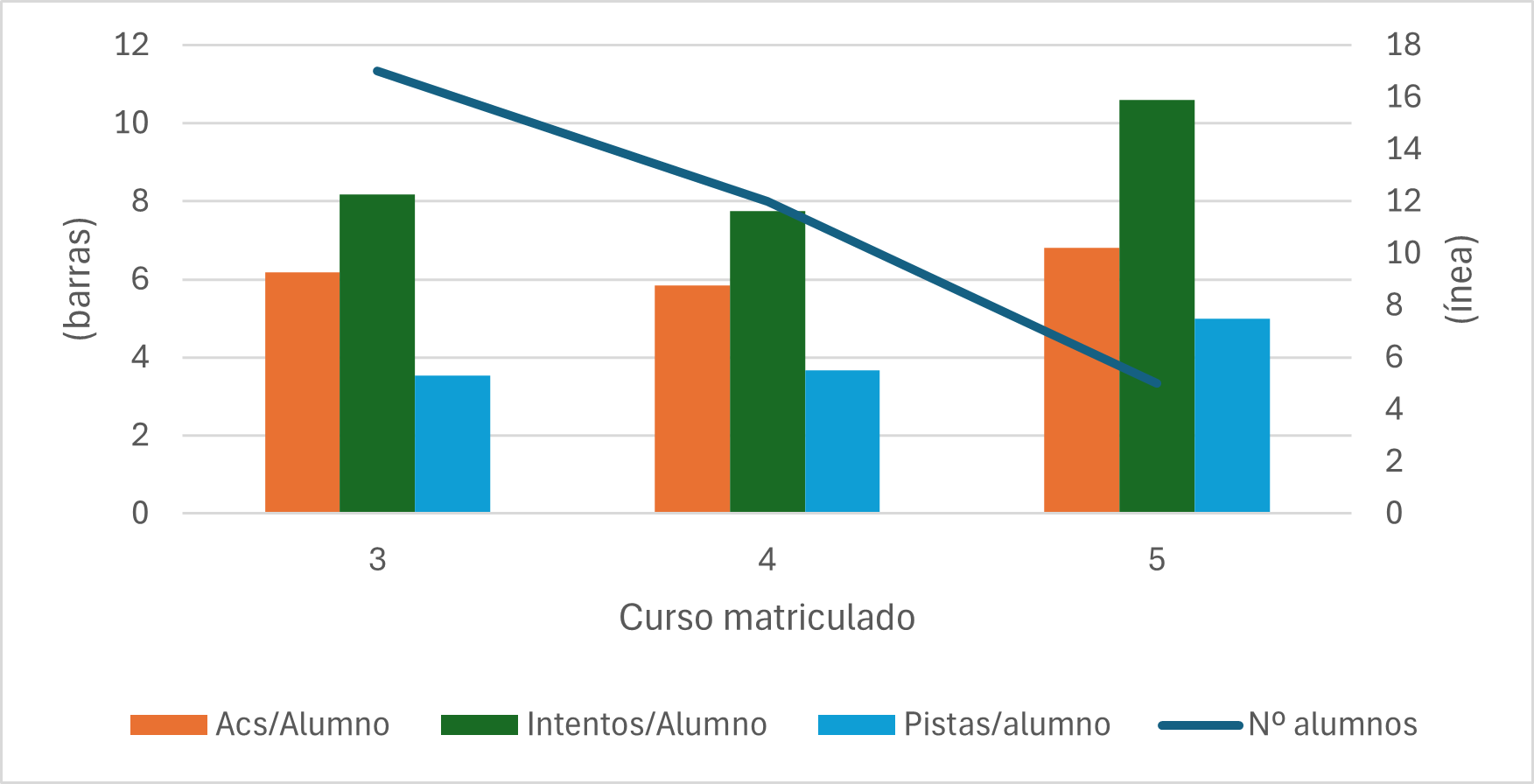}
    }
    \vspace{1em}
    \caption{Resultados analíticos del concurso e información detallada por curso matriculado\label{fig:resultados_concurso_agrupado}.}
\end{figure*}

La figura~\ref{fig:resultados_concurso_agrupado} resume los resultados de ambos concursos. En la figura~\ref{fig:resultados_problemas} se detalla el número de problemas resueltos por cada participante, junto con los intentos realizados y las pistas solicitadas. Se aprecia que, en los tres indicadores –especialmente en la cantidad de concursantes que se quedan en cada problema–, los datos siguen una distribución normal, lo que confirma un equilibrio adecuado en la dificultad: ni tan bajos como para sugerir un concurso excesivamente difícil, ni tan altos que impliquen una falta de reto. Destaca el problema 7 (la mitad más uno), que marca un notable salto en dificultad y actúa como barrera para los concursantes de menor nivel. Asimismo, el problema 5 es el que registra la mayor proporción de envíos incorrectos, ya que introduce aritmética, condicionales y bucles en \bash, cuya compleja sintaxis permite errores silenciosos. En este caso, un bucle con una condición de parada mal escrita funcionaba en el caso público pero generaba un bucle infinito en los privados, detectado por el sistema. Para futuras ediciones, se tendrá en cuenta este particular y se ofrecerá información más detallada sobre la sintaxis en la documentación y las pistas.

El número de envíos incorrectos es moderado pues la dificultad del concurso radica en dominar la sintaxis del lenguaje y encontrar en el manual las opciones adecuadas para los comandos, en contraste con los problemas de algoritmia donde el reto está en elaborar una solución genérica correcta en casos límite. Por ello, es habitual que un script comprobado con los casos de ejemplo funcione para los casos privados, cuya razón de ser es evitar la aceptación de soluciones hechas completamente {\em ad-hoc} a los ejemplos.

La figura~\ref{fig:resultado_curso} presenta un análisis detallado según el curso de matrícula de los estudiantes. En líneas generales, el rendimiento, medido en función del número de envíos correctos, es homogéneo entre los participantes. No se observa una gran diferencia entre los alumnos de tercer curso, que cursan su primera asignatura en la que utilizan Linux (SO), y los de cuarto y quinto curso, que ya han completado una asignatura adicional (ASOR). Esto sugiere que usos avanzados de \bash, como los que se aprenden a lo largo del concurso, no están incluidos en el plan docente, lo que corrobora la utilidad del concurso, que complementa los conocimientos reglados. Un análisis similar se refleja en el número de pistas solicitadas. Aunque los alumnos de quinto curso han resuelto ligeramente más problemas, lo que se traduce en un mayor número de intentos, este dato resulta poco concluyente debido a que son solo 2 participantes de los grupos de Doble Grado (grupos más reducidos), frente a 17 de tercer curso y 12 de cuarto curso.

Finalmente, es importante destacar que, aunque el concurso fue publicitado a todos los estudiantes de la facultad, no participaron alumnos de primer y segundo curso. Esto podría deberse a que el primer contacto con \bash se produce en una asignatura de tercer curso (Sistemas Operativos), lo que puede resultar intimidante para quienes aún no han tenido ningún contacto con el lenguaje. Además, los estudiantes de tercer, cuarto y quinto curso en Sistemas Operativos y Ampliación de Sistemas Operativos y Redes recibían un incentivo en forma de un ligero aumento de nota basado en su rendimiento en el concurso, beneficio que no tenían los alumnos de primero y segundo.

\subsection{Sobre la importancia y utilidad de \bash en el entorno académico y profesional}\label{sec:encuestas_pre}

La mayoría de los participantes en la encuesta indicaron que su principal motivación para participar en el concurso es la obtención de puntos adicionales en las asignaturas de Sistemas Operativos y Ampliación de Sistemas Operativos y Redes. En particular, 10 estudiantes seleccionaron la opción ``Obtener reconocimiento académico'' en la pregunta A6, mientras que solo 5 manifestaron que su objetivo principal era ``Aprender más sobre \bash''. No obstante, la percepción general sobre la utilidad de \bash es positiva, ya que los resultados de la pregunta A2 muestran que 11 participantes consideran que su utilidad es ``media'' y otros 11 la califican como ``alta'', lo que sugiere un reconocimiento de su relevancia tanto en el ámbito académico como profesional.

Esta valoración refuerza la hipótesis inicial de este estudio sobre la necesidad de fomentar el aprendizaje de \bash y, en general, el uso de la línea de comandos entre los estudiantes. De hecho, los propios encuestados reconocen la importancia de ampliar la formación en este ámbito, tanto dentro de las asignaturas del plan de estudios (pregunta A4, donde 8 estudiantes señalaron una necesidad ``media'', 12 una necesidad ``alta'' y 6 una necesidad muy ``alta'') como a través de actividades extracurriculares, tales como seminarios y cursos externos (pregunta A5, con 7 votos para necesidad ``media'', 10 para necesidad ``alta'' y 9 para necesidad ``muy alta'').

Estos resultados están respaldados no solo por la percepción del profesorado, sino también por la propia autovaloración de los estudiantes. En la pregunta A1, 19 de los encuestados indicaron poseer ``conocimientos medios'' de \bash, lo que sugiere que, si bien están familiarizados con el entorno de línea de comandos, también son conscientes de la necesidad de seguir profundizando en su uso.

Finalmente, a pesar de reconocer la importancia y utilidad de \bash, los resultados de la encuesta revelan que el lenguaje de scripting más utilizado actualmente por los estudiantes es Python, con 20 de los 26 encuestados eligiéndolo en la pregunta A7, seguido de JavaScript. Cabe destacar que, al igual que \bash, Python no forma parte del currículo formal del grado; sin embargo, su entorno de desarrollo más amigable, junto con la disponibilidad de herramientas, documentación y una comunidad de apoyo más amplia, lo han convertido en una opción preferida para los estudiantes que están aprendiendo a programar.

\subsection{Sobre la percepción de estudiantes hacia el concurso}\label{sec:encuestas_post}

En la pregunta A6, un total de 16 estudiantes indicaron que su motivación para participar en el concurso no estaba relacionada con ``Obtener reconocimiento académico'', sino con otras razones como ``Aprender más sobre \bash'' (5 votos), ``Divertirme y desafiarme a mí mismo/a'' (6 votos) o un interés general por la programación y los concursos de programación. Estos resultados sugieren que el diseño del concurso logra captar el interés de los estudiantes gracias a su componente interactivo y desafiante. Sin embargo, se podría explorar la posibilidad de fortalecer aún más su vínculo con competencias laborales y académicas, con el fin de maximizar su impacto en la formación profesional de los participantes.

Asimismo, la mayoría de los estudiantes consideran que el concurso es una herramienta útil y efectiva para demostrar sus conocimientos. En la pregunta B1, 7 encuestados indicaron que su utilidad es ``media'', mientras que 9 la calificaron como ``alta'' y 6 como ``muy alta''. Además, los participantes valoraron positivamente tanto la variedad de los ejercicios como el nivel de dificultad de los mismos (pregunta B3), con 10 votos a favor del ``nivel de dificultad de los ejercicios'' y 9 votos en apoyo de la ``variedad de los ejercicios''.

A pesar de que durante el concurso los estudiantes tenían acceso a páginas de manual y a una ficha resumen con las principales construcciones de \bash, los resultados reflejan que este material no fue percibido como una ayuda suficiente. En la pregunta B4, 9 estudiantes indicaron que la ``documentación aportada'' podría mejorarse en futuras ediciones. Esto podría deberse a que la documentación proporcionada es de carácter técnico, por lo que se plantea la posibilidad de complementarla con códigos de ejemplo más didácticos en versiones posteriores del concurso.

Por otro lado, un aspecto destacado por los participantes fue la necesidad de mejorar el ``hardware/software disponible''. En particular, las críticas posteriores al concurso señalaron la imposibilidad de utilizar herramientas gráficas, como Visual Studio Code, para conectarse de forma remota y editar los scripts. No obstante, esta restricción fue implementada de manera intencionada, con el objetivo de resaltar la dependencia que muchos estudiantes tienen de entornos gráficos con autocompletado. Si bien estas herramientas facilitan el trabajo, a menudo ocultan los conocimientos y técnicas subyacentes, lo que podría limitar el aprendizaje profundo de \bash y la línea de comandos.

\section{Conclusiones}\label{sec:conclusions}

En este trabajo, se ha presentado un sistema para la realización de concursos interactivos orientados al aprendizaje de Bash. La plataforma ha demostrado ser funcional y efectiva, permitiendo la gestión estructurada de los desafíos, la incorporación de nuevos problemas y su reutilización en futuras ediciones, lo que garantiza su sostenibilidad a largo plazo.

Los resultados del estudio indican que la gamificación mediante concursos constituye una estrategia didáctica eficaz para fomentar el aprendizaje de Bash. La mayoría de los participantes valoraron positivamente la experiencia, destacando su carácter dinámico y motivador. Asimismo, los datos recopilados reflejan que los estudiantes consideran necesario reforzar la enseñanza de Bash dentro del grado, dado que lo perciben como una herramienta de gran utilidad para su desarrollo académico y profesional.

No obstante, se ha identificado como área de mejora la documentación y formación previa proporcionada a los participantes. Los resultados sugieren que los materiales de referencia disponibles antes y durante el concurso no fueron suficientes para afrontar determinados desafíos, lo que pone de manifiesto la necesidad de complementar estos recursos en futuras ediciones.

\bibliographystyle{Jornadas}
\bibliography{biblio}

\begin{thebibliography}{10}

\bibitem{no-gui}
Ira Goldstein,
\newblock ``{What! No GUI?--Teaching a Text Based Command Line Oriented Introduction to Computer Science Course},''
\newblock {\em Information Systems Education Journal}, vol. 17, no. 1, pp. 40--48, 2019.

\bibitem{teaching_bash}
Victoria Plemakova,
\newblock ``Teaching advanced Bash scripting using semi-automated assessment environment,''
\newblock M.S. thesis, University of Tartu. Institute of Computer Science, 2016.

\bibitem{gamificacion-gen-z}
Hadeel~Mohammed Jawad and Samir Tout,
\newblock ``Gamifying Computer Science Education for Z Generation,''
\newblock {\em Information}, vol. 12, no. 11, 2021.

\bibitem{programming-contests}
L.H. Gonz\'{a}lez~Guerra and V.M. de~la Cueva,
\newblock ``The use of programming contests + positive feedback to inspire computer science students to improve their problem-solving skills,''
\newblock in {\em ICERI2023 Proceedings}, 2023, pp. 3393--3400.

\bibitem{competitions-motivation}
Youry Khmelevsky and Ken Chidlow,
\newblock ``Students Programming Competitions as an Educational Tool and a Motivational Incentive to Students'' 2021,
\newblock \url{https://arxiv.org/abs/2105.15136}.

\bibitem{deterding2011game}
Sebastian Deterding, Dan Dixon, Rilla Khaled, and Lennart Nacke,
\newblock ``From game design elements to gamefulness: defining" gamification",''
\newblock in {\em Proceedings of the 15th international academic MindTrek conference: Envisioning future media environments}, 2011, pp. 9--15.

\bibitem{DuolingoReview}
Mitchell Shortt, Shantanu Tilak, Irina Kuznetcova, Bethany Martens, and Babatunde Akinkuolie,
\newblock ``Gamification in mobile-assisted language learning: a systematic review of Duolingo literature from public release of 2012 to early 2020,''
\newblock {\em Computer Assisted Language Learning}, vol. 36, no. 3, pp. 517--554, 2023.

\bibitem{KahootReview}
Alf~Inge Wang and Rabail Tahir,
\newblock ``The effect of using Kahoot! for learning--A literature review,''
\newblock {\em Computers \& Education}, vol. 149, pp. 103818, 2020.

\bibitem{revilla2008UVa}
Miguel~A Revilla, Shahriar Manzoor, and Rujia Liu,
\newblock ``Competitive learning in informatics: The UVa online judge experience,''
\newblock {\em Olympiads in informatics}, vol. 2, no. 10, pp. 131--148, 2008.

\bibitem{mirzayanov2020codeforces}
Mike Mirzayanov, Oksana Pavlova, Pavel MAVRIN, Roman Melnikov, Andrew Plotnikov, Vladimir Parfenov, and Andrew Stankevich,
\newblock ``Codeforces as an educational platform for learning programming in digitalization,''
\newblock {\em Olympiads in Informatics}, vol. 14, no. 133-142, pp. 14, 2020.

\bibitem{Kattis}
Emma Enström, Gunnar Kreitz, Fredrik Niemelä, Pehr Söderman, and Viggo Kann,
\newblock ``Five years with kattis — Using an automated assessment system in teaching,''
\newblock in {\em 2011 Frontiers in Education Conference (FIE)}, 2011, pp. T3J--1--T3J--6.

\bibitem{gomez2017aceptaelreto}
Pedro~Pablo G{\'o}mez-Mart{\'\i}n and Marco~Antonio G{\'o}mez-Mart{\'\i}n,
\newblock ``¡Acepta el reto!: juez online para docencia en espa{\~n}ol,''
\newblock 2017,
\newblock \url{https://aceptaelreto.com}.

\bibitem{SWERC}
``ICPC Southwestern Europe Regional Contest (SWERC)'' \url{https://swerc.eu/2025/}.

\bibitem{Ada-Byron}
``Concurso Universitario de Programación AdaByron'' \url{https://ada-byron.es/2025/}.

\bibitem{Programame}
``Concurso de programación para Ciclos Formativos: ProgramaMe'' \url{https://programame.com/2025/reg/}.

\bibitem{OIE}
``Olimpiada Informática Española (OIE)'' \url{https://olimpiada-informatica.org/}.

\bibitem{Leal2003Mooshak}
José~Paulo Leal and Fernando Silva,
\newblock ``Mooshak: a Web-based multi-site programming contest system,''
\newblock {\em Software: Practice and Experience}, vol. 33, no. 6, pp. 567--581, 2003.

\bibitem{DOMjudge}
``Sistema de automzatización para concursos de programación DOMjudge'' \url{https://www.domjudge.org/}.

\bibitem{concursoSARTECO}
``Concurso de programación paralela de la Sociedad de Arquitectura y Tecnología de Computadores (SARTECO)'' \url{http://luna.inf.um.es/2018/}.

\bibitem{CMDChallenge}
``Web de retos en Bash CMD Challenge'' \url{https://cmdchallenge.com/}.

\bibitem{ExercismBash}
``Itinerio de aprendizaje Bash de la web Exercism'' \url{https://exercism.org/tracks/bash}.

\end{thebibliography}

\end{document}